\documentclass[pra,reprint,superscriptaddress]{revtex4-2}
\usepackage[utf8]{inputenc}
\usepackage[T1]{fontenc}
\usepackage{float}
\usepackage{caption}
\usepackage{subcaption}
\usepackage{graphicx}
\usepackage{blindtext}
\usepackage{xcolor}
\usepackage{todonotes}
\usepackage{lineno}
\usepackage{mathtools, cuted}
\usepackage{amsmath}
\usepackage{amssymb}
\usepackage{tikz}
\usetikzlibrary{positioning,arrows}
\usetikzlibrary{decorations.pathmorphing}
\usetikzlibrary{decorations.markings}
\usepackage{siunitx}
\definecolor{revblue}{cmyk}{0.95,0.85,0.01,0.04}

{\catcode `\@=11 \global\let\AddToReset=\@addtoreset}
\AddToReset{equation}{section}

\newcounter{mnotecount}[section]
\renewcommand{\themnotecount}{\thesection.\arabic{mnotecount}}
\newcommand{\mnotex}[1]
{\protect{\stepcounter{mnotecount}}$^{\mbox{\footnotesize
$
\bullet$\themnotecount}}$ \marginpar{
\raggedright\tiny\em
$\!\!\!\!\!\!\,\bullet$\themnotecount: #1} }

\newcommand{\vect}[1]{\boldsymbol{#1}}

\usepackage{chngcntr}
\counterwithout{equation}{section}
\begin{document}

\title{Requirements on Quantum Superpositions of Macro-Objects for Sensing Neutrinos}

\author{Eva Kilian}
\email{eva.kilian.18@ucl.ac.uk}
\affiliation{%
 Department of Physics \& Astronomy, University College London, WC1E 6BT London, UK 
}

\author{Marko Toro\v{s}}
\affiliation{%
School of Physics \& Astronomy, University of Glasgow, Glasgow, G12 8QQ, UK
}
\author{Frank F. Deppisch}
\affiliation{%
 Department of Physics \& Astronomy, University College London, WC1E 6BT London, UK 
}
\author{Ruben Saakyan}
\affiliation{%
 Department of Physics \& Astronomy, University College London, WC1E 6BT London, UK 
}
\author{Sougato Bose}
\affiliation{%
 Department of Physics \& Astronomy, University College London, WC1E 6BT London, UK 
}
%

\date{\today}

\begin{abstract}
We examine a macroscopic system in a quantum superposition of two spatially separated localized states as a detector for a stream of weakly interacting relativistic particles. We do this using the explicit example of neutrinos with \si{MeV}-scale energy scattering from a solid object via neutral-current neutrino-nucleus scattering. Presuming the (anti-)neutrino source to be a nuclear fission reactor, we utilize the estimated flux and coherent elastic neutrino-nucleus cross section to constrain the spatial separation $\Delta{x}$ and describe the temporal evolution of the sensing system. Particularly, we find that a potentially measurable relative phase between quantum superposed components is obtained for a single gram scale mass placed in a superposition of spatial components separated by $10^{-14}$~\si{m} under sufficient cooling and background suppression.  
\end{abstract}
\maketitle

\section{Introduction} Despite extensive scientific efforts, neutrinos still pose a puzzling enigma, decades after they were first observed experimentally~\cite{reines_neutrino_1956}. While it may be known that neutrinos interact only through the weak and the gravitational forces, many of the questions on the very nature of these particles remain unanswered to this day. As all other fermions in the Standard Model, neutrinos were formerly assumed to be representable by Dirac spinors and additionally thought to be massless. However, oscillation experiments have shown that particles produced in a particular, well-defined flavour eigenstate can, after having travelled a sufficiently long distance, with a certain probability be detected in a different flavour state~\cite{fukuda_evidence_1998}. A consequence of these findings is that neutrinos do have mass and that their flavor eigenstates are different from their mass eigenstates. As a result of being massive, neutrinos could be either Dirac or Majorana particles and it is currently unknown which of the two they are. Present day oscillation experiments enable estimates of mass squared differences of the three neutrino mass eigenstates, but they are neither capable of measuring absolute neutrino masses, nor have so far been able to shed light on the possibility of the existence of further sterile neutrinos. Limits on the latter two quantities do however exist. The vast multitude of unknowns provides a motivation to seek novel methods to detect neutrinos, especially to examine if the detector size can be reduced. In this work, we will introduce one such approach aiming to study reactor anti-neutrinos with energies of a few \si{MeV} through their {\em momentum transfer} in scattering from a macroscopic system placed in a quantum superposition of distinct centre of mass positions -- with the momentum transfer appearing as a relative phase between the components of the superposition.\\

\begin{figure}[!t]
    \centering
    \includegraphics[width=.5\textwidth]{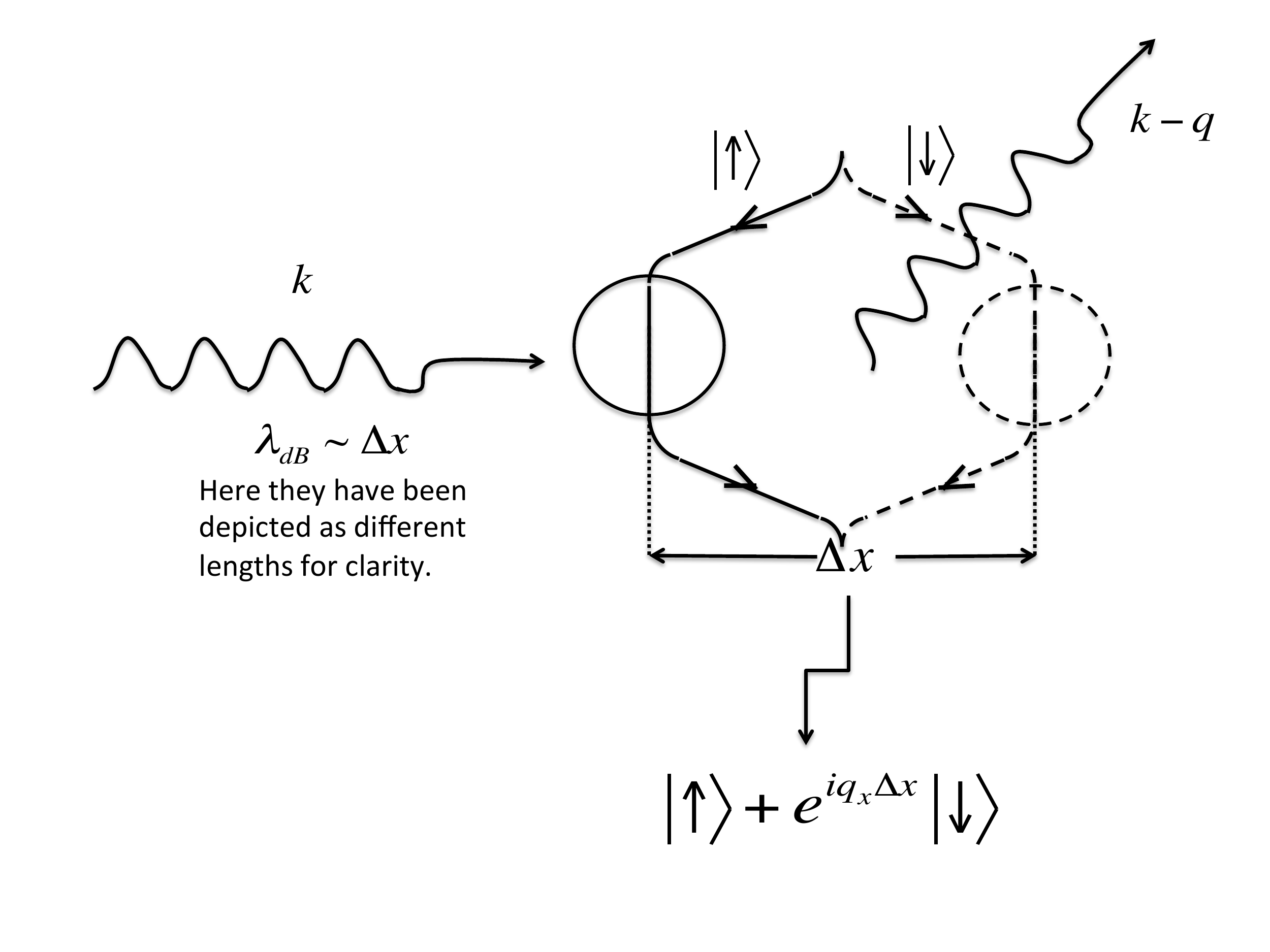}
    \caption{The working mechanism of a system detecting the momentum recoil of a crystal due to the scattering of a particle from it through a phase between two components of a superposition. In a general Stern-Gerlach interferometer, the phase difference between two spatially separated components end up as a phase difference between two spin states and can be measured as a phase difference between spin components.}
    \label{Scatt}
\end{figure}

The field of matter-wave interferometry in which a large mass, such as a solid object or crystal of several atoms, goes to a quantum superposition of being ``here'' and ``there''  is an emerging area still in development, with several nascent ideas. For an inexhaustive list see  \cite{bose1999scheme, armour2002entanglement,marshall2003towards,PhysRevA.84.052121,romero2011large,bateman2014near,scala2013matter,yin2013large,PhysRevLett.117.143003,pedernales2020motional,margalit2021realization,wood2022spin,marshman2021large}.  These developments have so far been primarily steered by the aim to extend the boundaries of quantum mechanics {\em empirically} to larger objects, as the quantum behaviour the centre of mass (COM) of a sufficiently macroscopic masses remains untested.  Unlike in the case of interference experiments with several atoms from a cold source, for example a Bose-Einstein condensate,  where each atom goes one way or another,  here {\em all} atoms of the crystal go one way together or {\em all} atoms of the crystal go the other way together.  On  the experimental side these developments are stimulated, on the one hand, by the demonstration of quantum superpositions of the COM of large molecules consisting of up to $\sim 2000$ atoms~\cite{fein2019quantum} and on the other hand, by the achievement of cooling of the COM of much larger masses such as $10^{-17}$ kg nanocrystals~\cite{magrini_real-time_2021,delic2020cooling,  tebbenjohanns2020motional, tebbenjohanns_quantum_2021} and 10 kg masses  \cite{whittle2021approaching} close to their quantum mechanical ground states.  While this may still indicate a significant gap between what has been demonstrated,  and what needs to be achieved in order to realise experiments  with the COM of large masses in a superposition,  there are the above well formulated schemes and conditions which could be adapted.  

Quantum superpositions of the COM of large masses can have a great potential as a sensor~\cite{rademacher_quantum_2020}. It has already been theoretically demonstrated that they can measure tiny gravitational effects including the detection of low frequency gravitational waves \cite{marshman_mesoscopic_2020}, and, ultimately, even evidence the quantum nature of gravity~\cite{bose_spin_2017,marletto2017gravitationally, marshman2020locality,  bose2022mechanism} or be able to test new forces \cite{barker2022entanglement} and the weak Equivalence principle in a quantum regime \cite{bose2022entanglement}. Once a quantum superposition of a large mass being in two positions is produced,  external forces cause a relative phase shift between the two components of this superposition. Hence, we propose that it might also be possible to detect tiny momentum transfers due to the scattering of weakly interacting particles from such superpositions.  In particlular, recently the approach of detecting particles beyond the standard model via the momentum recoil of several levitated nano-objects in localized (classical) states has become topical \cite{monteiro2020search,  afek-2022-coherent}.  It is natural thus to ask whether quantum superpositions can aid further.  We study this herewith using neutrinos as an example, importantly in a regime in which they scatter coherently from entire nuclei as that significantly enhances the cross section. \\

Predicted more than 40 years ago in 1974~\cite{freedman_coherent_1974} and recently observed experimentally~\cite{akimov_observation_2017}, coherent elastic scattering of neutrinos from nuclei (CE$\nu$NS) is the dominant scattering channel for incoming neutrino energies $E_\nu \ll 100 ~\si{MeV}$. The scattering coherence manifests in an enhancement of the cross-section, which scales with $N^2$,  with $N$ being the number of neutrons in the nucleus. Up until recently, it has however been of great difficulty to detect such neutral current events, in part due to the large detector volumes (enough nuclei) and low energy thresholds required to detect the \si{keV} to sub-\si{keV} recoils of the nuclei.\\

In light of the advent of proposals for small-scale neutrino detectors~\cite{domcke_detection_2017,strauss_-cleus_2017} that aim to exploit the small recoil energies and relatively large CE$\nu$NS cross section associated with scattering events in the (sub-)\si{MeV} neutrino energy regime, we consider the suitability of matter-wave interferometric schemes for detecting such processes. We emphasise that in contrast to classical methods, our approach is based on the detection of neutrinos via massive quantum devices and hence exploits features which are inherently quantum. \\

\section{Detecting particulate matter by measuring a phase}
\label{sec:two}
The essense of the type of detector that we are considering here is given in Fig.\ref{Scatt}. A mass is prepared in a quantum superposition of spatially distinct states. An ideal way to generate such a spatial superposition is by employing a Stern-Gerlach type interferometric scheme with a single spin embedded in a mass.  An archetypal example is the Nitrogen-Vacancy (NV) point defect in a diamond crystal, which carries a spin-1 (made of two electrons) \cite{bar2013solid}.  However,  any other point defect with an electronic spin in any other crystal \cite{niethammer2019coherent} or a single dopant atom with an unpaired electronic  spin implanted in a solid, as for example, used in certain designs of solid state quantum computers  \cite{jakob2022deterministic} will serve our purpose.  It is understood that the atom carrying this spin is tightly bound to the rest of the crystal.  Also we have to ensure that there is this single spin which can couple to external magnetic fields strongly (i.e. with the strength of a Bohr magneton)  while there may be unpaired nuclear spins on various atoms in the crystal which couple with the much lower strength of nuclear magnton.  While detailed studies can be found in Refs.\cite{scala2013matter,yin2013large,PhysRevLett.117.143003,pedernales2020motional,margalit2021realization,wood2022spin,marshman2021large,keil2021stern}, here we present only a schematic description.  The spin is initially prepared in a quantum superposition \mbox{$\frac{1}{\sqrt{2}}(|\uparrow\rangle_s + |\downarrow\rangle_s)$} and the mass bearing the spin is subjected to an inhomogeneous magnetic field. This couples its spin and motional degrees of freedom: for $|\uparrow\rangle_s$ spin state, the mass accelerates to the left, and for the $|\downarrow\rangle_s$ spin state, the mass accelerates to the right. They can be brought to a halt at a given superposition size $\Delta x$ by flipping the spins at appropriate times. The resulting entangled state between spin and centre of mass degree of freedom of the mass thus generated is
\begin{align}\label{eq:initial-state}
    |\Psi_0\rangle_{S}&=
\frac{1}{\sqrt{2}}\big(|\uparrow\rangle_{s}|\bar{x}_0\rangle_{C}+|\downarrow\rangle_{s}|\bar{x}_1\rangle_{C}\big),
\end{align}
where $|\bar{x}_0\rangle_C$ and $|\bar{x}_1\rangle_C$ respectively refer to both parts of the superposition, each one describing the center-of-mass motion of our test mass (crystal), while the subscript $S$ denotes the combined system of spin and crystal. The states $|\bar{x}_j\rangle$ are to be understood as Gaussian wavepackets localized around the position $(\bar{x}_j,0,0)$, and momentum $p \approx 0$. The superposition size is $\Delta x = |\bar{x}_1\rangle-|\bar{x}_0\rangle$. For simple notation, we treat the macroscopic mass as a single particle and write the COM state of the macroscopic mass as
\begin{align}\label{eq:initial-second-Gaussian}
    |\bar{x}_j\rangle_C= \Bigl(\frac{1}{\sigma_c\sqrt{2 \pi}}\Bigr)^{\frac{1}{2}} \int_{-\infty}^{\infty} dx~e^{-\frac{(x-\bar{x}_j)^2}{2\sigma_c^2}}\Psi^{\dagger}_C(x)|0\rangle,
\end{align}
where $\Psi^{\dagger}_C(x)$ creates the whole crystal at position $(x, 0, 0)$ and $|0\rangle$ is the vacuum state. Fundamentally, $|\bar{x}_j\rangle$ is a many-particle state. Hence $\Psi^{\dagger}(x)_C$ is equal to a product of proton, neutron and electron creation operators at positions locked to, and distributed around, the COM position $(x,0,0)$. 

Let us now find out what happens to the above Gaussian state $|\bar{x}_j\rangle$ when a momentum $\vect{q}$ is transferred to it. To model this, we assume momentum creation and annihilation operators of the mass to be $b^{\dagger}_{\vect{k}}$ and $b_{\vect{k}}$ respectively,
\begin{align}\label{eq:initial}
   &(\int d^3\vect{k}~ b^{\dagger}_{\vect{k}+\vect{q}}b_{\vect{k}}) \int_{-\infty}^{\infty} dx~e^{-\frac{(x-\bar{x}_j)^2}{2\sigma_c^2}}\Psi^{\dagger}(x)|0\rangle \nonumber \\
&\propto \int_{-\infty}^{\infty} dx~e^{-\frac{(x-\bar{x}_j)^2}{2\sigma_c^2}} \int d^3 \vect{k}~b^{\dagger}_{\vect{k}+\vect{q}}b_{\vect{k}} \int d^3\vect{k}^{'} e^{i k_x^{'} x} b^{\dagger}_{\vect{k}^{'}} |0\rangle \nonumber \\
&=\int_{-\infty}^{\infty} dx~e^{-\frac{(x-\bar{x}_j)^2}{2\sigma_c^2}} e^{-i q_x x} \int d^3 (\vect{k}+\vect{q})~e^{i (k_x+q_x) x}b^{\dagger}_{\vect{k}+\vect{q}}|0\rangle \nonumber \\
&=\int_{-\infty}^{\infty} dx~e^{-\frac{(x-\bar{x}_j)^2}{2\sigma_c^2}} e^{-i q_x x} \Psi^{\dagger}(x)|0\rangle \nonumber \\
&=e^{-iq_x \bar{x}_j} e^{-\frac{{q_x}^2\sigma_c^2}{2}} \int_{-\infty}^{\infty} dx~e^{-\frac{(x-\bar{x}_j+iq_x\sigma_c^2)^2}{2\sigma_c^2}}\Psi^{\dagger}(x)|0\rangle.
\end{align}
Considering the special case when the width of the Gaussian wavepacket is much smaller than the length scale of the transferred momentum $\sigma_c\lesssim 1/q_x$, the state of the Gaussian after the momentum transfer is 
\begin{align}\label{eq:gaussian-kick}
e^{-iq_x \bar{x}_j}  \int_{-\infty}^{\infty} dx~e^{-\frac{(x-\bar{x}_j)^2}{2\sigma_c^2}}\Psi^{\dagger}(x)|0\rangle
= e^{-iq_x \bar{x}_j}|\bar{x}_j\rangle,
\end{align}
implying that the initial state of a superposed mass and a scattering particle evolves as 
\begin{align}\label{eq:superpose-phase}
 |\Psi_0\rangle_S|\vect{p}\rangle_B\rightarrow |\Psi_{q_x}\rangle_S|\vect{p}-\vect{q}\rangle_B 
, \end{align}
where we have introduced the label $B$ to denote the bath environment comprised of the scattering particle(s) and the subscript $S$ to refer to the superposed target mass. The state \mbox{$|\Psi_{q_x}\rangle_S$} is given by 
\begin{align}\label{eq:superpose-phase-explicit}
|\Psi_{q_x}\rangle_S=\frac{1}{\sqrt{2}}\big(|\uparrow\rangle|\bar{x}_0\rangle+e^{-iq_x \Delta x}|\downarrow\rangle|\bar{x}_1\rangle\big),
\end{align}
where the difference in center-of-mass position is $\Delta x= \bar{x}_1-\bar{x}_0$.
A particle scattering from the above state and transferring a momentum $\vect{q}$ to it, could be detected as a phase difference of $ \frac{q_x \Delta x}{\hbar}$ (restoring the $\hbar$) between the components of the superposition, where $q_x$ is the $x$-component of the momentum transfer. In a typical Stern-Gerlach interferometry experiment, the two components $|\bar{x}_0\rangle$ and $|\bar{x}_1\rangle$ are brought back to completely overlap with each other by reversing the process which created them, as shown in Fig. \ref{Scatt} so that the spin state becomes $|\psi_{q_x}\rangle_s=\frac{1}{\sqrt{2}}\big(|\uparrow\rangle_s+e^{-iq_x \Delta x}|\downarrow\rangle_s\big)$. Thus the phase is measurable purely from the off-diagonal component $\langle \uparrow| \rho_s |\downarrow\rangle$ of the density matrix $\rho_s=|\psi_{q_x}\rangle\langle \psi_{q_x}|_s$ of the spin. In general, the scattering neutrinos will scatter with a distribution of momenta, so that there will be a mixed state density matrix, while the process itself will lead to a decoherence as well. We will therefore treat the evolution in terms of open quantum systems techniques, deriving a master equation to be followed by the density matrix.

\section{Coherent Elastic Neutrino-Nucleus Scattering Cross Section}
In this section, we discuss the CE$\nu$NS cross section in the regime of reactor anti-neutrinos, since the calculation of the matrix element is directly related to the interaction Hamiltonian. For this purpose, we outline the description of coherent elastic scattering of a neutrino with incident four momentum $p_i=(E_\nu,\mathbf{p}_i)$ by a nucleus of mass $m_{\text{nucl}}$ and incident four momentum $k_i=(m_{\text{nucl}},0)$, restricting ourselves to neutral current processes for the reason of simplicity. An illustration of the process is depicted in figure $\ref{fig:elastic_scatt}$, where we have assumed the comparably heavy nucleus to be at rest and the final momenta to be of the form $p_f=(E_{\nu,f},\mathbf{p}_f)$ and $k_f=(E_{n,f},\mathbf{k}_f)$.

\tikzset{
particle/.style={draw=blue, postaction={decorate},
    decoration={markings,mark=at position .5 with {\arrow[draw=blue]{>}}}},
gluon/.style={decorate, draw=black,
    decoration={coil,amplitude=4pt, segment length=5pt}}
 }

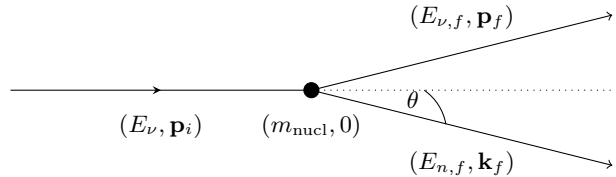
\begin{figure}[H]
\centering
    \begin{tikzpicture}[decoration={markings, 
    mark= at position 0.5 with {\arrow{stealth}}}] 
    \draw[postaction={decorate}] (0,0) -- (4,0) node[below, pos=0.5, yshift=-0.2cm] {$(E_\nu,\mathbf{p}_i)$};
    \draw[dotted] (4,0) -- node [below]{}++(4,0);
    \draw[->] (4,0)-- node[above,yshift=0.2cm] {$(E_{\nu,f},\mathbf{p}_f)$} ++(4,1);
    \draw[->] (4,0)-- node[below,yshift=-0.2cm] {$(E_{n,f},\mathbf{k}_f)$} ++(4,-1);
    \draw[fill=black] (4,0) circle (1mm) node[below, yshift=-0.2cm] {$(m_{\text{nucl}},0)$};
   \draw (5.5,0) arc (55:10:.7);
   \draw (5.35,.1) node[below] {$\theta$};
    \end{tikzpicture}
    \caption{Elastic scattering illustration with the scattering angle denoted as $\theta$, the initial momenta $p_i$, $k_i$ and the final momenta $p_f$, $k_f$.}
    \label{fig:elastic_scatt}
\end{figure}

 The incoming and outgoing four momenta can be reformulated in terms of the incoming neutrino energy $E_\nu$, the kinetic energy transferred due to scattering $T$ and the nucleus mass $m_{\text{nucl}}$,

\begin{align}
    p_i&=(E_\nu,\mathbf{p}_i),\\
    k_i&=(m_{\text{nucl}},0),\\
    p_f&=(E_\nu-T,\mathbf{p}_f),\\
    k_f&=(m_{\text{nucl}}+T,\mathbf{k}_f).
\end{align}

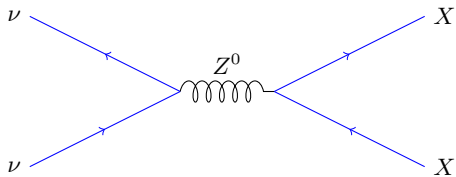
\begin{figure}[h!]
    \centering
   \begin{tikzpicture}[node distance=1cm and 2cm]
\coordinate[label=left:$\nu$] (e1);
\coordinate[below right=of e1] (aux1);
\coordinate[right=1.25cm of aux1] (aux2);
\coordinate[above right=of aux2,label=right:$X$] (e2);
\coordinate[below left=of aux1,label=left:$\nu$] (e3);
\coordinate[below right=of aux2,label=right:$X$] (e4);

\draw[particle] (e3) -- (aux1);
\draw[particle] (aux1) -- (e1);
\draw[particle] (e4) -- (aux2);
\draw[particle] (aux2) -- (e2);
\draw[gluon] (aux1) -- node[label=above:$Z^0$] {} (aux2);
\end{tikzpicture}
    \caption{Feynman diagram for neutral current coherent elastic $\nu$-nucleus scattering, where time runs vertically. The label $\nu$ refers to the neutrino whereas the symbol $X$ represents the nucleus.}
    \label{fig:nun_scatt}
\end{figure}

The tree-level Feynman diagram representing such a process is shown in figure \ref{fig:nun_scatt}.

In order to recover the cross section, the standard approach is to determine the scattering amplitude $\mathcal{M}_{ss'}$ using the neutral current interaction Lagrangian

\begin{align}\label{eq:lagr}\mathcal{L}=\frac{G_F}{\sqrt{2}}J^\mu_{\,\nu} J_{\mu,\text{nucl}},
\end{align}

and the relevant Feynman rules \cite{lindner_coherent_2017}. The quantities described by $J^\mu_{\nu}$ and $J_{\mu, \text{nucl}}$ denote the neutral currents of neutrino and nucleus. Neutral fermionic currents are given by 

\begin{align}
    J^\mu_{f}=\sum_i\overline{\psi}_{f_i}\gamma^\mu(g_V^{f_i}-g_A^{f_i}\gamma^5)\psi_{f_i}
\end{align}

where $\psi_f$ denote the respective fermionic fields. Omitting the sum and inserting its couplings $g_V^\nu=1/2$ and $g_A^\nu=1/2$, the neutrino neutral current is

\begin{align}
    J^\mu_{\nu}&=\overline{\psi}_{\nu}\gamma^\mu\frac{1}{2}(1-\gamma^5)\psi_{\nu}.
\end{align}

It is a general feature of the weak interaction that its exchange bosons couple only to left-handed neutrinos or right-handed anti-neutrinos. For the nucleus neutral current, we choose a description in terms of its weak coupling constant and the four-momenta $k_i$, $k_f$,

\begin{align}
    J_{\mu, nucl}&=\frac{Q_W}{2}F(q^2)(k_i+k_f)_\mu,
\end{align}

where $Q_W$ refers to the weak charge and $F(q)$ to the form factor dependent on the exchanged momentum $q$. Equation (\ref{eq:lagr}) can now be used to recover the Feynman amplitude $\mathcal{M}$, which is then formulated as follows

\begin{align}\label{eq:ampl}
     i\mathcal{M}_{ss'}&=i\frac{G_FQ_WF(q^2)g^\nu_L}{\sqrt{2}}\nonumber\\
     &\times\overline{u}^{s'}(p_f)\gamma^{\mu}(1-\gamma^5)u^s(p_i)(k_i+k_f)_\mu.
\end{align}

In the laboratory frame, the differential cross section with respect to the transferred kinetic energy is related to the absolute square of the matrix element as $\frac{d\sigma}{dT}=\frac{\sum_{ss'}|i\mathcal{M}_{ss'}|^2}{32\pi m_{\text{nucl}}E_{\nu}}$ and hence given by~\cite{scholberg_coherent_2015}

\begin{align}
    \frac{d\sigma}{dT}=\frac{G_F^2Q_W^2|F(q^2)|^2m_{\text{nucl}}}{4\pi}\left(1-\frac{T}{E_\nu}-\frac{m_{\text{nucl}}T}{2E_\nu^2}\right).\label{eq:cross_sect}
\end{align}

The differential cross section for the scattering of anti-neutrinos is derived analogously. Assuming a scattering angle $\theta$ between the initial neutrino and final nucleus momenta, the kinetic energy $T$ can be expressed as follows

\begin{align}
    T=\frac{2m_{\text{nucl}}E_\nu^2\cos^2\theta}{(m_{\text{nucl}}+E_\nu)^2-E_\nu^2\cos^2\theta}.
\end{align}

The maximum kinetic energy transfer is then obtained for $\theta$=0. The coherent elastic neutrino-nucleus cross section can be recovered from Eq.~(\ref{eq:cross_sect}), following integration over kinetic energy. In all further calculations, we have approximated the form factor as $F(q)\sim 1$. This is appropriate solely for the energy range considered. It should be noted that the cross section increases not only with neutrino energy, but also with increasing number of neutrons $N$ in the target nucleus. The latter dependency becomes evident through a closer inspection of the weak charge

\begin{align}
     Q_W=((1-4\sin^2\theta_W)Z+N).
\end{align}

For the purpose of a real-world experiment, this means that the target material does play an important role in the detectability of any scattering phase effect from neutrino scattering on nuclei.

\section{Quantum Open Systems and the Born-Markov Approximation}
We consider the scenario of a heavy nucleus in the presence of a (fermionic) bath of neutrinos scattering from it. As we have no control over a specific scattered neutrino, but nonetheless wish to describe the evolution of a comparably heavy nucleus, we resort to an open quantum systems approach. Following the description in~\cite{agarwal_quantum_2012}, the total Hamiltonian of a weakly interacting system and bath is
\begin{equation}
    H=H_S+H_B+H_{SB}
\end{equation}
where $H_{SB}$ refers to the interaction between $S$ and $B$. In order to derive the master equation of the system, it is of convenience to write the interaction Hamiltonian as
\begin{equation}
    H_{SB}=\sum_\alpha S_\alpha B_\alpha
\end{equation}
with $S_\alpha$ and $B_\alpha$ denoting the system and bath operators.  We will hence have to map the respective parts of our non-standard interaction Hamiltonian for coherent elastic neutrino-nucleus scattering onto system and bath operators. The summation will become an integral over all relevant external three momenta. It shall also be noted that we impose hermiticity on $H_{SB}$. Taking the bath correlation time to be much shorter than the timescale over which our system evolves, we resort to a description in the Born-Markov approximation. The master equation for the system is then governed by
\begin{equation}
\begin{aligned}
\label{eq:sys_evol}
    \frac{d\rho_S}{dt}&=-\frac{i}{\hbar}[H_S,\rho_S]-\Bigl\{\int^\infty_0 d\tau \sum_{\alpha\beta}C_{\alpha\beta}(-\tau)\\
    &\times[S_\alpha S_\beta(-\tau)\rho_S
    -S_\beta(-\tau)\rho_SS_\alpha]+\text{H.c.}\Bigr\},
\end{aligned}
\end{equation}
where 
\begin{equation}
\begin{aligned}
    C_{\alpha\beta}=\frac{1}{\hbar^2}Tr[\rho_BB_\alpha B_\beta(-\tau)]
\end{aligned}
\end{equation}
is the bath correlation function and the time dependence of the operators $S_\beta$ and $B_\beta$ is determined by
\begin{align}
    S_\beta(-\tau)=e^{-\frac{iH_S\tau}{\hbar}}S_\beta e^{\frac{iH_S\tau}{\hbar}},\\
    B_\beta(-\tau)=e^{-\frac{iH_B\tau}{\hbar}}B_\beta e^{\frac{iH_B\tau}{\hbar}}.
\end{align}
As we work in natural units, we set $\hbar=c=1$ for now.

\subsection{Quantum Master Equation for Neutrino-Nucleus Scattering}
The Lagrangian density for the interaction of neutrino $\nu$ and nucleus $n$ is is given by Eq.~($\ref{eq:lagr}$). Hence, the interaction Hamiltonian $H_{n,\nu}$ corresponds to 
\begin{equation}
\begin{aligned}
    H_{n,\nu}&=-\int d^3x\mathcal{L}(x)\\
    &=-\frac{G_F}{\sqrt{2}}\int d^3xJ^{\mu}_{\nu}(x)J_{\mu,\text{nucl}}(x) =H_{SB}.
\end{aligned}
\end{equation}
Choosing to rewrite the neutrino current in second quantised form, we observe that
\begin{equation}
\begin{aligned}
    J^\mu_\nu(x)&=\overline{\psi}_\nu(x)\frac{1}{2}\gamma^\mu(1-\gamma^5)\psi_\nu(x)\\
    &=\int \frac{d^3p_id^3p_f}{(2\pi)^6\sqrt{4E_{p_i}E_{p_f}}}a^{\dagger}_{p_f}a_{p_i}e^{-i(p_i-p_f)x}\\
    &\times\overline{u}^2(p_f)\frac{1}{2}\gamma^\mu(1-\gamma^5)u^2(p_i),
\end{aligned}
\end{equation}

so that our Hamiltonian is
\begin{equation}
\begin{aligned}
    H_{n,\nu}&=-\frac{G_FQ_W F(q)}{2\sqrt{2}}\int  d^3x\\
    &\times\frac{d^3p_id^3p_fd^3k_id^3k_f}{(2\pi)^{12}\sqrt{16E_{p_i}E_{p_f}E_{k_i}E_{k_f}}}\\ &\times e^{-i(k_i+p_i-k_f-p_f)x}
     a^{\dagger}_{p_f}a_{p_i}\\
     &\times\overline{u}^{s'}(p_f)\gamma^{\mu}(1-\gamma^5)u^s(p_i)c^{\dagger}_{k_f}c_{k_i}(k_f+k_i)_\mu.
\end{aligned}
\end{equation}
Note that although here we have generically used $k_i,  k_f$ to label 3-momenta,  $(k_f+k_i)_\mu$ stand for 4-momenta.
Noting that the only $x$-dependence is now in the exponential, we can perfom the spatial integration. Further to this, the factor $F(q)\sim1$ in our scenario and can therefore be neglected,
\begin{equation}
\begin{aligned}
    H_{n,\nu}&=-\frac{(2\pi)^3G_FQ_W}{2\sqrt{2}}\int \frac{d^3p_id^3p_fd^3k_id^3k_f}{(2\pi)^{12}\sqrt{16E_{p_i}E_{p_f}E_{k_i}E_{k_f}}}\\ &\times\delta^3(k_i+p_i-k_f-p_f)a^{\dagger}_{p_f}a_{p_i}\\
    &\times \overline{u}^{s'}(p_f)\gamma^{\mu}(1-\gamma^5)u^s(p_i)c^{\dagger}_{k_f}c_{k_i}(k_f+k_i)_\mu\\
&=\frac{(2\pi)^3G_FQ_W}{2\sqrt{2}}\int \frac{d^3p_id^3p_fd^3k_id^3k_f}{(2\pi)^{12}16E_{p_i}E_{p_f}E_{k_i}E_{k_f}}\\ &\times\overline{u}^{s'}(p_f)\gamma^{\mu}(1-\gamma^5)u^s(p_i)(k_f+k_i)_\mu\\
    &\times\delta^3(k_i+p_i-k_f-p_f)|p_f\rangle\langle p_i|\otimes|k_f\rangle\langle k_i|,
\end{aligned}
\label{eq:hamiltonian}
\end{equation}
with one-particle states $|p,s\rangle=\sqrt{2E_p}a^{s\dagger}_p|0\rangle$. Using the matrix element for the neutrino-nucleus scattering $\mathcal{M}_{p_i,k_i,p_f,k_f}=\frac{G_FQ_W}{2\sqrt{2}}\overline{u}^{s'}(p_f)\gamma^{\mu}(1-\gamma^5)u^s(p_i)(k_f+k_i)_\mu$ and by computing the integral over the final nucleus momentum $k_f$ we get
\begin{equation}
\begin{aligned}
    \hat{H}_{n,\nu}&=\int \frac{d^3p_i d^3p_fd^3k_id^3k_f}{(2\pi)^{9} 2E_{p_i}2E_{p_f}2E_{k_i}2E_{k_f}}\mathcal{M}_{p_i,k_i,p_f,k_f} \\
    &\delta^3(k_i+p_i-k_f-p_f)|p_f\rangle\langle p_i|\otimes|k_f\rangle\langle k_i|\\
    &=\int \frac{d^3p_i d^3p_fd^3k_i\mathcal{M}_{p_i,k_i,p_f,k_i+p_i-p_f}}{(2\pi)^{9} 2E_{p_i}2E_{p_f}2E_{k_i}2E_{k_i+p_i-p_f}}\\
    &|p_f\rangle\langle p_i|\otimes|k_i+p_i-p_f\rangle\langle k_i|\\
    &=\int \frac{d^3p_i d^3p_fd^3k_i\mathcal{M}_{p_i,k_i,p_f,k_i+p_i-p_f}}{(2\pi)^{9} 2E_{p_i}2E_{p_f}2E_{k_i}2E_{k_i+p_i-p_f}}\\
    &|p_f\rangle\langle p_i|\otimes e^{i(p_i-p_f)\hat{x}}|k_i\rangle\langle k_i|
\end{aligned}
\end{equation}
where we have used the fact that the momentum state $|k_i+p_i-p_f\rangle$ can be rewritten in a suitable manner.
Now we make a crucial simplifying approximation: we assume the nucleus mass to be sufficiently large compared to its momenta $m_{\text{nucl}}\gg |k|$. Being inside the crystal which is stationary, we treat the nucleus as being effectively at rest so that the expression of the matrix element implies $\mathcal{M}_{p_i,k_i,p_f,k_f} \sim \mathcal{M}_{p_i,0,p_f,0}$. Let us write the nucleus integrals explicitly, so that $d\mu_\nu$ now only comprises the neutrino momentum integrals.
\begin{equation}
\begin{aligned}
    \hat{H}_{n,\nu}&=\int d\mu_\nu \int d^3k_i\frac{\mathcal{M}_{p_i,0,p_f,0}}{(2\pi)^3(2E_{k_i}2E_{k_i+p_i-p_f})}\\
    &\times |p_f\rangle\langle p_i|\otimes e^{i(p_i-p_f)\hat{x}}|k_i\rangle\langle k_i|
\end{aligned}
\end{equation}
For a very heavy nucleus, let us further assume that we can approximate the quantity $2E_{k_i}2E_{k_i+p_i-p_f}\sim 4m_{\text{nucl}}E_{k_i}$. Seeing as the difference in the neutrino momenta is small and the energy of the nucleus is predominantly dependent on its heavy mass, we take this to be a reasonable justification for the model at hand. As a consequence, we observe that the Hamiltonian reduces further, for we are now able to use the definition for the one-particle state identity resolution.
\begin{equation}
\begin{aligned}
    \hat{H}_{n,\nu}=\int d\mu_\nu\frac{\mathcal{M}_{p_i,p_f}}{2m_{\text{nucl}}} |p_f\rangle\langle p_i|\otimes e^{i(p_i-p_f)\hat{x}}\mathbb{I}
\end{aligned}
\end{equation}
Next, we associate the operators of the bath with the integrals over the amplitudes, whereas the rest of the above Hamiltonian is kept in the system operators.
In our case, we identify 
\begin{equation}
\begin{aligned}
    S_\alpha&=\frac{1}{2m_{\text{nucl}}}e^{i(p_i-p_f)\hat{x}}\mathbb{I}, ~~
    B_\alpha=\mathcal{M}_{p_i,p_f}|p_f\rangle\langle p_i|,
\end{aligned}
\end{equation}
 and we pack all dependencies on the Feynman amplitudes into the bath operators. Further, we argue that the COM of the crystal (to which the nucleus belongs) is trapped in a very low frequency trap, so that the time evolution of the operator $S_\beta$ can be neglected.
In order to time evolve the neutrino state, we assume the neutrino rest mass to be negligible with respect to its total energy. We further neglect flavor oscillations, which is a reasonable assumption over the short distances of $d\sim 20\si{m}$ we consider, so that the neutrino bath has the free Hamiltonian 
\begin{equation}
\begin{aligned}
    H_B=\int \frac{d^3p}{(2\pi)^3}E_{p}a^\dagger_pa_p.
\end{aligned}
\end{equation}

At first we will consider a single neutrino scattering from the nucleus, thus we need to find the bath correlation function for a single neutrino bath. In order to do that, we need to use an incoming neutrino in a sufficiently momentum localized state $|\psi\rangle$ normalised to $\langle\psi|\psi\rangle=1$. This is achieved by starting with a generic Gaussian initial momentum state
\begin{equation}
    \begin{aligned}
        |\psi\rangle&=\int \frac{d^3p}{(2\pi)^3\sqrt{2E_p}}\psi(p)|p\rangle\\
        &=\int \frac{d^3p}{(2\pi)^3\sqrt{2E_p}}\frac{(2\pi)^{3/2}}{(2\pi\sigma^2)^{3/4}}e^{-\frac{(p-p_0)^2}{4\sigma^2}}|p\rangle.
    \end{aligned}
\end{equation}
We substitute $\tilde{\sigma}=\sqrt{2}\sigma$, such that our state is properly normalised to $\langle\psi|\psi\rangle=1$ and regard it in the limit of a very narrow wavefunction.
\begin{equation}
    \begin{aligned}
        \lim_{\tilde{\sigma}\rightarrow 0}|\psi\rangle&=\lim_{\tilde{\sigma}\rightarrow 0}\int\frac{d^3p\,(4\pi\tilde{\sigma}^2)^{3/4}e^{-\frac{(p-p_0)^2}{2\tilde{\sigma}^2}}}{(2\pi)^{3/2}\sqrt{2E_p}(2\pi\tilde{\sigma}^2)^{3/2}}|p\rangle\\
        &=\epsilon\int\frac{d^3p}{\sqrt{2E_p}}(\tilde{\sigma})^{3/2}\delta^3(p-p_0)|p\rangle\\
        &=\frac{\epsilon\tilde{\sigma}^{3/2}}{\sqrt{2E_{p_0}}}|p_0\rangle
    \end{aligned}
\end{equation}

Here, $\epsilon$ represents the collected numerical factors. We will use this simplified form in the last step above for our subsequent computations. Thus,  
\begin{equation}
\begin{aligned}
    C_{\alpha,\beta}&=Tr[\rho_\nu B_\alpha B_{\beta}(-\tau)]\\
    &=Tr[|\psi\rangle\langle\psi|\mathcal{M}_{p_i,p_f}\mathcal{M}^*_{p'_f,p'_i}e^{-i(E_{p'_i}-E_{p'_f})\tau}|p_f\rangle\\
    &\times\langle p_i|p'_f\rangle\langle p'_i|]\\
    &=(2\pi)^3\mathcal{M}_{p_i,p_f}\mathcal{M}^*_{p'_f,p'_i}2E_{p_i}\delta^3(p_i-p_f')\\
    &\times \langle \psi|\psi\rangle\langle\psi|e^{-i(E_{p'_i}-E_{p'_f})\tau}|p_f\rangle\langle p'_i|\psi\rangle \\
&=(2\pi)^3\mathcal{M}_{p_i,p_f}\mathcal{M}^*_{p'_f,p'_i}2E_{p_i}\delta^3(p_i-p_f')\\
    &\times\langle\psi|e^{-i(E_{p'_i}-E_{p'_f})\tau}|p_f\rangle\langle p'_i|\psi\rangle\\
    &=(2\pi)^9\epsilon^2\tilde{\sigma}^{3}\mathcal{M}_{p_i,p_f}\mathcal{M}^*_{p'_f,p'_i}2E_{p_i}\delta^3(p_i-p_f')\\
    &\times e^{-i(E_{p'_i}-E_{p'_f})\tau}2E_{p'_i}\delta^3(p_0-p_f)\delta^3(p'_i-p_0).
\end{aligned}
\end{equation}
For $d\rho_S/dt$ this results in the second term of Eq.~(\ref{eq:sys_evol}) given by 
\begin{equation}
\begin{aligned}
    \frac{d\rho_S}{dt}&=-\frac{\epsilon^2\tilde{\sigma}^{3}}{4m_{\text{nucl}}^2}\int d\tau\int \frac{d^3p_id^3p_fd^3p'_id^3p'_f}{(2\pi)^{3}16E_{p_i}E_{p_f}E_{p'_i}E_{p'_f}}\\
    &\times 4E_{p_i}E_{p'_i}\mathcal{M}_{p_i,p_f}\mathcal{M}^*_{p'_f,p'_i}\delta^3(p_i-p_f')\\
    &\times e^{-i(E_{p'_i}-E_{p'_f})\tau}\delta^3(p_0-p_f)\delta^3(p_i'-p_0)\\
    &\times \{-e^{i(p'_i-p'_f)\hat{x}}\rho_S e^{i(p_i-p_f)\hat{x}}\\
    &+e^{i(p_i-p_f)\hat{x}}e^{i(p'_i-p'_f)\hat{x}}\rho_S+\text{H.c.}\}\\
    &=-\frac{\epsilon^2\tilde{\sigma}^{3}}{4m_{\text{nucl}}^2}\int d\tau\int \frac{d^3p_fd^3p'_id^3p'_f}{(2\pi)^{3}4E_{p_f}E_{p'_f}}\\
    &\times\mathcal{M}_{p'_f,p_f}\mathcal{M}^*_{p'_f,p'_i}e^{-i(E_{p'_i}-E_{p'_f})\tau}\delta^3(p_0-p_f)\\
    &\times\delta^3(p_i'-p_0)\{-e^{i(p'_i-p'_f)\hat{x}}\rho_S e^{i(p'_f-p_f)\hat{x}}\\
    &+e^{i(p'_f-p_f)\hat{x}}e^{i(p'_i-p'_f)\hat{x}}\rho_S+\text{H.c.}\}\\
    &=-\frac{\epsilon^2\tilde{\sigma}^{3}}{8m_{\text{nucl}}^2E_{p_0}}\int d\tau\frac{d^3p'_f}{(2\pi)^{3}2E_{p'_f}}|\mathcal{M}_{p_0,p_f'}|^2 \\
    &\times e^{-i(E_{p_0}-E_{p'_f})\tau}\{-e^{i(p_0-p'_f)\hat{x}}\rho_S e^{i(p'_f-p_0)\hat{x}}\\
    &+e^{i(p'_f-p_0)\hat{x}}e^{i(p_0-p'_f)\hat{x}}\rho_S+\text{H.c.}\}.
\end{aligned}
\end{equation}
Lastly, we have
\begin{equation}
\begin{aligned}
    \frac{d\rho_S}{dt}&=-\frac{\epsilon^2\tilde{\sigma}^{3}}{64\pi^{2}m_{\text{nucl}}^2E_{p_0}}\int \frac{d^3p'_f}{E_{p'_f}}|\mathcal{M}_{p_0,p_f'}|^2\\
    &\times \delta(E_{p_0}-E_{p'_f})\{-e^{i(p_0-p'_f)\hat{x}}\rho_S e^{i(p'_f-p_0)\hat{x}}\\
    &+e^{i(p'_f-p_0)\hat{x}}e^{i(p_0-p'_f)\hat{x}}\rho_S+c.c.\}.
\end{aligned}
\end{equation}
It is always possible to find a suitable parameterisation of the momenta $p_0$ and $p_f'$ in terms of an energy and appropriate angles. Therefore $\mathcal{M}_{p_0,p_f'}=\mathcal{M}(E_{p_f'}, \Omega)\equiv\mathcal{M}(\Omega)$. Setting $\alpha=\frac{\epsilon^2\tilde{\sigma}^{3}}{64\pi^{2}m_{\text{nucl}}^2}$ for brevity, we are able to write the factor before the curly brackets as
\begin{equation}
\begin{aligned}
    \Gamma&=-\alpha\int \frac{E_{p'_f}^2dE'_{p_f}d\Omega}{E_{p_0}E_{p'_f}}|\mathcal{M}(E_{p_f'}, \Omega)|^2 \delta(E_{p_0}-E_{p'_f})\\
    &=-\frac{\epsilon^2\tilde{\sigma}^{3}}{64\pi^{2}m_{\text{nucl}}^2}\int d\Omega|\mathcal{M}(\Omega)|^2.
\end{aligned}
\end{equation}

It shall be noted that the factor \mbox{$\epsilon^2\tilde{\sigma}^{3}_p=2^{-\frac{3}{2}}(2\pi\tilde{\sigma}^{2}_x)^{-\frac{3}{2}}=(2\pi\sigma^{2}_x)^{-\frac{3}{2}}=V_x^{-1}$}, which is essentially the volume of a Gaussian times a factor. Seeing as \mbox{$\sigma_x=(\int x^2|\psi(x)|^2dx)^{1/2}$} for a conventionally normalised $\psi(x)$, our normalised Gaussian in position space yields the recovered prefactor. The prefactor can be interpreted as the expectation value of finding a particle within the volume element $V_x$, which implies that number of particles per unit volume $n=1/V_x$. The neutrino's velocity $|p_\nu|/E_\nu\sim c= 1$. Hence, its flux is given by 
\begin{equation}
    \begin{aligned}
        F=nc=n,
    \end{aligned}
\end{equation}
where $n$ is the number of particles per unit volume. We have normalized the single neutrino wavefunction to 1 particle per unit volume. Thus we have, in terms of the flux of 1 particle, $F_1$, the evolution of the reduced density matrix of the nucleus as given by
\begin{equation}
\begin{aligned}
    \frac{d\rho_S}{dt}&=-\frac{F_1}{64\pi^{2}m_{\text{nucl}}^2}\int d\Omega|\mathcal{M}(\Omega)|^2\\
    &\times\{-e^{i(\Delta(E_0,\Omega))\hat{x}}\rho_S e^{-i(\Delta(E_0,\Omega))\hat{x}}+\rho_S+c.c.\},
\end{aligned}
\end{equation}
and hence
\begin{equation}
    \begin{aligned}
        \langle x|\overset{\cdot}{\rho}_S|y\rangle&=-\frac{2F_1}{64\pi^{2}m_{\text{nucl}}^2}\int d\Omega|\mathcal{M}(\Omega)|^2\\
    &\times\{-e^{i(\Delta(E_0,\Omega))(x-y)}+1\}\langle x |\rho_S|y\rangle.
    \end{aligned}
    \label{eq:one_nu_scatt}
\end{equation}
In writing the above, we have implicitly assumed that the system, i.e. the nucleus, has a negligible evolution due to its own Hamiltonian $H_S$ during the time scale of the experiment.\\
We are now in a position to compute the change in the density matrix of the centre of mass of the whole crystal which is comprised of multiple nuclei subject to a large flux of neutrinos from a reactor. It shall be noted that our approximations yield a result which is qualitatively very close to the form of Gallis-Fleming~\cite{gallis_environmental_1990} for non-relativistic particles scattering from a mass. Moreover, Eq.~(\ref{eq:one_nu_scatt}) has a very intuitive interpretation, with $\frac{F_1}{64\pi^{2}m_{\text{nucl}}^2} d\Omega|\mathcal{M}(\Omega)|^2$ being the incident flux multiplied by the scattering cross section for a solid angle $d\Omega$. It is thus the rate of scattering in a given solid angle $d\Omega$. Each scattering direction $d\Omega$ imparts a different momentum to the nucleus, which is given by the operator $e^{i(\Delta(E_0,\Omega))\hat{x}}$.

\subsection{Calculation of the Relative Phase between Superposed Components of a Crystal and its Detection}

In order for our formalism to apply to a whole crystal in the form of a bulk material consisting of multiple nuclei scaling as $N_{\text{Atoms}}$ and a flux of incoming neutrinos with a spectral distribution of energies $S(E)$, Eq.~(\ref{eq:one_nu_scatt}) will have to be modified to

\begin{equation}
\label{Final}
\begin{aligned}
    \frac{d\rho_S}{dt}&=-\frac{2F_1\cdot N_{\text{Atoms}}}{64\pi^{2}m_{\text{nucl}}^2}\int dES(E)\int d\Omega|\mathcal{M}(\Omega)|^2\\
    &\{-e^{i(\Delta(E_0,\Omega))\hat{x}}\rho_S e^{-i(\Delta(E_0,\Omega))\hat{x}}+\rho_S+c.c.\}.
\end{aligned}
\end{equation}

\begin{table}[t!]
    \centering
    \begin{tabular}{|c|c|}
        \hline G$_F$&1.1664$\cdot10^{-11}[\si{MeV^{-2}}]$\\\hline
         u&931.5$[\si{MeV\cdot c^{-2}}]$\\\hline
        $m_{\text{nucl}}$~\cite{lunney_recent_2003}&$(Z+N)u-0.00054858Z\cdot u+$\\
         &$(14.4381Z^{2.39}+1.55468\cdot10^{-6}Z^{5.35})10^{-6}$\\\hline
         Flux&$1.7\cdot 10^{13}[\si{s\cdot cm^{-2}}]$\\\hline
         $\Delta x$&$10^{-14}[\si{m}]$\\\hline
         S(E)&$\frac{1}{\sigma_E\sqrt{2\pi}}e^{-(E-E_0)^2/(2\sigma_E^2)}$\\\hline
         $\sigma_E$&0.75[\si{MeV}]\\\hline
         E$_0$&2.6[\si{MeV}]\\\hline
    \end{tabular}
    \caption{Constants and definitions. $G_F$ denotes the Fermi constant, $u$ the atomic mass unit, $m_{\text{nucl}}$ the mass of a nucleus. The flux listed is the projected neutrino flux at a distance of 20~$\si{m}$ from the source and $\Delta{x}$ refers the superposition size. The function $S(E)$ is a spectral distribution function over the energies $E$, with standard deviation $\sigma_E$ and mean energy $E_0$.}
    \label{tab:my_consts}
\end{table}

Anti-neutrino production rates for nuclear reactor sources are typically on the order of \mbox{$r\sim2\cdot10^{20}s^{-1}/\si{GW}_{th}$~\cite{hayes_reactor_2016,kim_detection_2016}}, with anti-neutrino energies ranging from $1-10
\si{MeV}$. Seeing as we assume our detectors to be placed at a distance d=$20~\si{m}$ to a 4.5 $\si{GW_{th}}$ nuclear fission reactor source, we obtain an estimated flux of

\begin{align}
    F_1=\frac{r_{4.5\,\si{GW}_{th}}}{4\pi d^2}\sim 1.7\cdot 10^{13}\si{cm^{-2}}\si{s^{-1}}.
\end{align}

As described earlier, the centre of mass $C$ of the crystal will be initialized in a joint state with with its spin $s$, in the motional superposition state \mbox{$|\Psi_0\rangle_{S}=\frac{1}{\sqrt{2}}\big(|\uparrow\rangle_{s}|\bar{x}_0\rangle_{C}+|\downarrow\rangle_{s}|\bar{x}_1\rangle_{C}\big)$}.

Labelling the orthormal states \mbox{$|\uparrow\rangle_{s}|\bar{x}_0\rangle_C$} and \mbox{$|\downarrow\rangle_{s}|\bar{x}_1\rangle_C$} with $|0\rangle$ and $|1\rangle$, respectively, for simplicity, we get the initial density matrix in the $\{|0\rangle,|1\rangle\}$ basis as

\begin{equation}
    \rho_0=\frac{1}{2}\begin{pmatrix}
    1&1\\
    1&1
    \end{pmatrix}.
\end{equation}

As we have discussed, for low momentum transfer with respect to the inverse of the width of the Gaussians $\bar{x}_0$ and $\bar{x}_1$, they can be treated effectively as position eigenstates $|\bar{x}_0\rangle$ and $|\bar{x}_1\rangle$ in the phase expression. Thus, the evolution of the density matrix in a time $\Delta t$, which we call the final density matrix $\rho_f$ is given by 
\begin{equation}
    \begin{aligned}
  \langle 0|  \rho_f|1\rangle &=\langle 0|  \rho_S(\Delta t)|1\rangle
\\ &=\langle 0|\rho_S(0)|1\rangle\\&-\frac{2F_1\cdot N_{\text{Atoms}}}{64\pi^{2}m_{\text{nucl}}^2}\int dES(E)\int d\Omega|\mathcal{M}(\Omega)|^2\\&
    \{-e^{i(\Delta(E_0,\Omega))(\bar{x}_0-\bar{x}_1)}+1\}\langle 0 |\rho_S(0)|1\rangle\Delta t.
    \end{aligned}
\end{equation}

As a result of the above evolution, we will obtain a final density matrix of the general form
\begin{equation}
    \rho_f=\begin{pmatrix}a&A e^{-i\varphi}\\Ae^{i\varphi}&b\end{pmatrix},
\label{rho}
\end{equation}
in terms of a phase $\varphi$ and an amplitude $A$. Unlike in the case of a simple phase acquisition, the amplitude accompanying the off diagonal term of the density matrix, where the phase is encoded, has also decayed because of the open systems treatment. In other words, we are averaging over all angles of scattering, which amounts to averaging over all momenta and incident energies. To extract the effect of the scattering, the centre of mass $C$ will be decoupled using the interferometry methods described in \cite{bose_spin_2017,marshman_mesoscopic_2020}. The relative phase acquired between $|\uparrow\rangle_{s}|\bar{x}_0\rangle_C$ and $|\downarrow\rangle_{s}|\bar{x}_1\rangle_{C}$ will appear between spin states $|\uparrow\rangle_{s}$ and $|\downarrow\rangle_{s}$ and be measured after suitable transformations between them. At the end of interferometry, $|\uparrow\rangle_{s}|\bar{x}_0\rangle_C$ and $|\downarrow\rangle_{s}|\bar{x}_1\rangle_{C}$ are mapped to spin states $|\uparrow\rangle_{s}$ and $|\downarrow\rangle_{s}$, respectively and we can continue to use $|0\rangle$ and $|1\rangle$ as our basis with the understanding that these now refer to the spin states $|\uparrow\rangle_{s}$ and $|\downarrow\rangle_{s}$ which are measured. Unitary operations on spin states typically comprise sending microwave pulses of appropriate frequencies to spin states~\cite{bar2013solid} differing in energies due to the Zeeman effect in an external magnetic field, hyperfine interactions or crystal field anisotropies. The entire toolbox of quantum computation is available and we will use two quantum operations, the Hadamard gate $H$ and the phase gate $S$ \cite{nielsen2002quantum}, on the spins before measuring the populations of $|\uparrow\rangle_{s}$ and $|\downarrow\rangle_{s}$.\\
After computing $\rho_f(x_1,x_2,t)$, upon applying a Hadamard transformation, we effectively rotate our bases from $|0\rangle\rightarrow\frac{1}{\sqrt{2}}(|0\rangle+|1\rangle)$ and $|1\rangle\rightarrow\frac{1}{\sqrt{2}}(|0\rangle-|1\rangle)$. The extraction of the phase then becomes a matter of calculating the probabilities of measuring $|0\rangle\langle 0|$ or $|1\rangle\langle 1|$. For the final density matrix a Hadamard transformation to the rotated basis yields

\begin{equation}
    H\rho_f H=\frac{1}{2}\begin{pmatrix}a+b+2A\cos{\varphi}&a-b+2iA\sin{\varphi}\\a-b-2iA\sin{\varphi}&a+b-2A\cos{\varphi}\end{pmatrix}
\end{equation}

and therefore, subtraction of the probabilities $p(0)-p(1)$ results in

\begin{equation}
    p(0)-p(1)=2A\cos{\varphi}.
\end{equation}

As it is sometimes more practical to express the phase for small arguments via the sine, a phase gate of the form

\begin{equation}
    S=\begin{pmatrix}1&0\\0&e^{i\pi/2}\end{pmatrix}
\end{equation}

is used before the Hadamard transformation to recover

\begin{equation}
    p_{S,H}(1)-p_{S,H}(0)=2A\sin{\varphi}.
\label{sinephase}
\end{equation}

\begin{figure}[!t]
 \includegraphics[width=.90\linewidth]{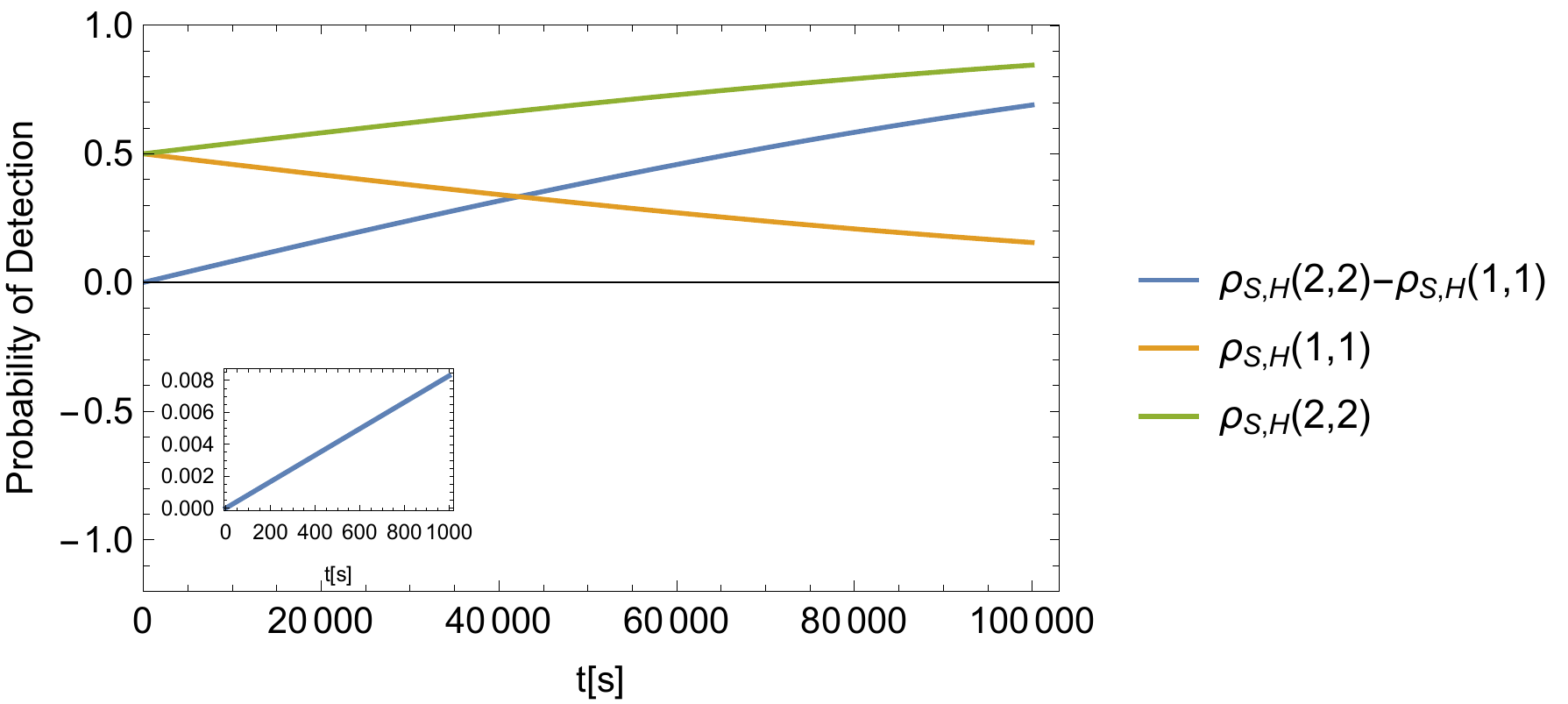}
\caption{Phase accumulation due to coherent neutrino-nucleus scattering from Bismuth. The figure depicts the normalised matrix elements of the nucleus density matrix after performing the operations of a phase gate and subsequent Hadamard on the sensing system. Most notably, the blue line shows the change resulting from the scattering in terms of the sine of the accumulated phase.}
\label{fig:PhaseAccumulation}
\end{figure}

For a crystal consisting of $N_{\text{Atoms}}=5\cdot10^{21}$ of the element $^{209}$Bi (crystal mass m $ \sim 1$ g) and the parameters as given in table \ref{tab:my_consts}, we observe phase accumulation and its amplitude decay in Fig. \ref{fig:PhaseAccumulation}. From the figure it is clear that at time $\sim 10^5$ s, a very significant phase difference between the components of the superposition with a significant amplitude is obtained for a superposition of size $\Delta x = |\bar{x}_0-\bar{x}_1| \sim 10^{-14}$ m. The separation was optimized and this order of magnitude was found to give a detectable phase with minimal damping at $\sim 10^5$ s. This time-scale corresponds to a $\sim \pi/3 \sim 1$ phase shift. About thrice this time corresponds to a $\pi$ phase shift as the phase growth is linear in time. For a $\pi$ phase shift, if we can ensure that there has been no other momenta imparting particle/effect, then this corresponds to the detection of one neutrino by our detector with 100$\%$ certainty (a "click" in our detctor) as such a phase is measured in a single shot by measuring the spin state in the $\{|+\rangle_s, |-\rangle_s\}$ basis with the outcome $|+\rangle_s$ corresponding to no neutrinos, and the outcome $|-\rangle_s$ corresponding to one neutrino. The chance of more than one neutrino scattering in the given time-scale is exceptionally small. Smaller non-zero phases ascertained at earlier times using Eq.(\ref{sinephase}) and multiple measurements to determine probabilities (i.e. repeating the procedure with the same detector or conducting measurements on an array of detectors) will detect the neutrino stream coming from the reactor, but will not be a single shot "click" detector. 


\section{Creation of Quantum Superpositions of Macroscopic Objects} 
\label{creation}
\begin{figure}[!t]
    \centering
    \includegraphics[width=.45\textwidth]{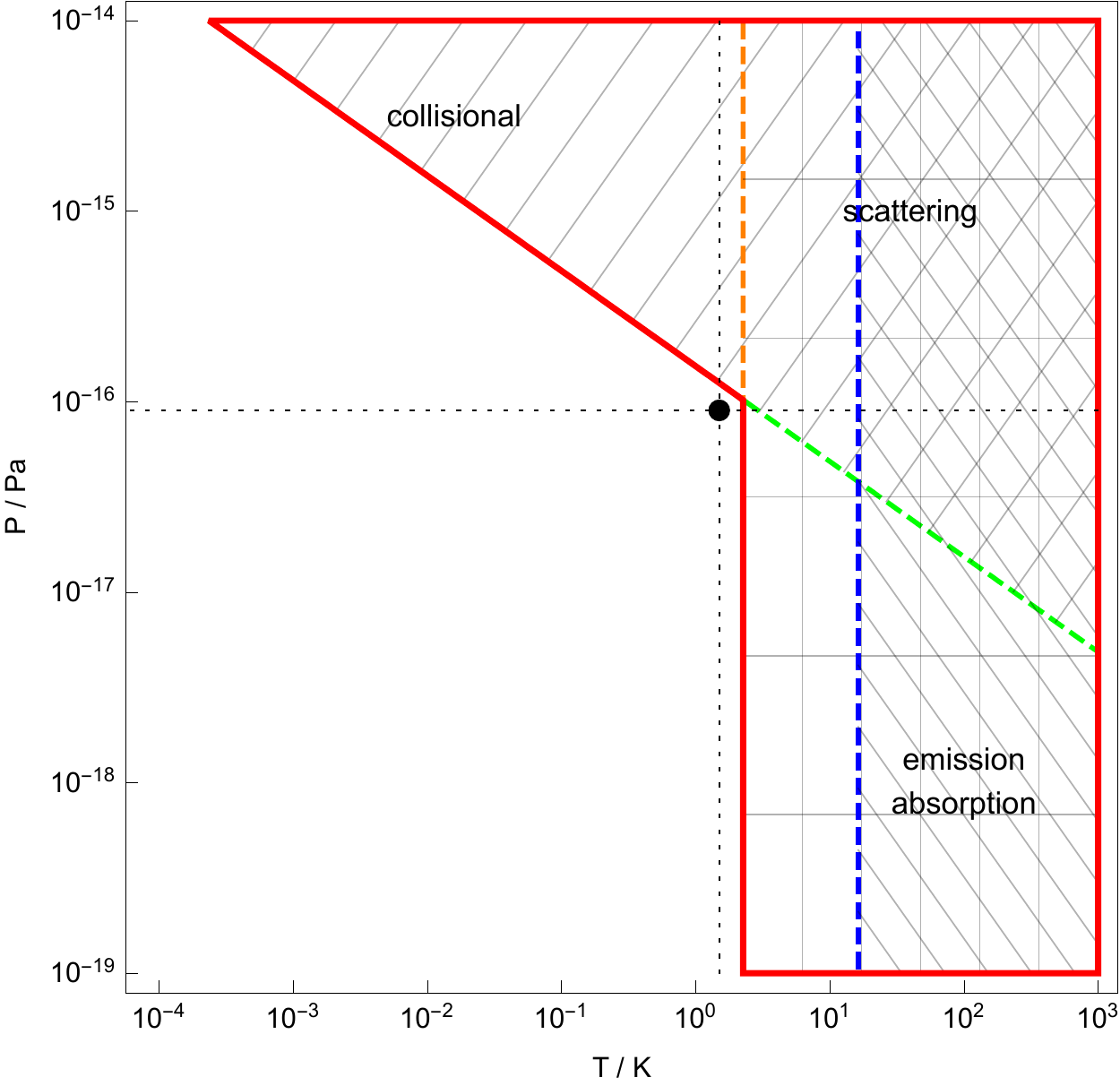}
    \caption{The diagram of the allowable region of pressure $P$ and temperature $T$ so that a $m\sim 1$ g crystal in a superposition of two positions separated by $\Delta x \sim 10^{-14}$ m can remain coherent for a time $t \sim 10^5$ s. The allowable region is unshaded. We can see that $P \sim 10^{-16}$ Pa and $T \sim 1$ K (the black dot) is an optimal point for our scheme.}
    \label{decoherence}
\end{figure}

We have three requirements for our setup. Firstly, the crystal should stay suspended against gravity for the duration of $\sim 10^5$ s of our experiment, although we will outline methods of reducing this time by resorting to a detector array rather than a single detector. Using whatever means, we have to trap the object in the vertical $z$ direction. This could be achieved via the well demonstrated mechanism of diamagnetic levitation which will balance the crystal against gravity. Once created, the quantum superposition of $m\sim 1$ g, $\Delta x \sim 10^{-14}$ m has to be kept coherent for $10^5$ s. This is a very long time, but the principal mechanisms of decoherence are known \cite{PhysRevA.84.052121}, namely the collisions with background gas (controlled by decreasing pressure) and black-body radiation emmision from the crystal (controlled by cooling the crystal internally). Fig.~\ref{decoherence} shows the requirements, with the unshaded region (outside the red bounded box) an allowed domain for coherence. It shows that pressures of $P \sim 10^{-16}$ Pa, already achieved in penning traps, and temperatures of $T\sim 1$K should suffice to retain the extremely long coherence for $10^5$ s. The effect of electromagnetic noise from the apparatus to create and probe the superposition is also of importance (analysed to some extent in~\cite{marshman_mesoscopic_2020}) and depends on the precise protocol, but essentially the exceptional stability of these sources, along with other proximal electromagnetic sensors will have to be used. For intertial noise, again, other sensors will have to be used to measure and take account of the noise. Alternatively, the detection can be done with two different materials in parallel, with the inertial noise being common. The explicit development of the above are beyond the scope of the current work, where we just want to highlight the possibility of neutrino detection.

Methods for creating superpositions of the form $|\Psi_0\rangle=\frac{1}{\sqrt{2}}\big(|\uparrow\rangle_s|\bar{x}_0\rangle_C+|\downarrow\rangle_s|\bar{x}_1\rangle_C\big)$ are still in development. Of course, our method will work for superpositions created by any means, even for superpositions without an ancillary spin system, such as $\frac{1}{\sqrt{2}}\big(|\bar{x}_0\rangle_C+|\bar{x}_1\rangle_C\big)$, as long as we have a mechanism to measure the relative phase between the components by bringing them together to interfere. It is just simpler for a superposition of the form $|\Psi_0\rangle=
\frac{1}{\sqrt{2}}\big(|\uparrow\rangle_s|\bar{x}_0\rangle_C+|\downarrow\rangle_s|\bar{x}_1\rangle_C\big)$, as the spin can be measured after the completion of interferometry to measure the phase. Here we only outline schematics rather than fully detailed schemes. One can use a mass with a single quantum spin$-1/2$ system embedded in it and subject it to magnetic field gradients. The $m\sim 1$ g crystal with an embedded spin is subjected to a $\frac{\partial B}{\partial x} \sim 10^6$ Tm$^{-1}$ (produced, for example, at a $\sim$ cm distance from the surface of a $\sim$ cm radius wire \cite{marshman_mesoscopic_2020} carrying $10^{13}$ A m$^{-2}$ current densities). When exposed to this magnetic field gradient for a time $t_{\text{acc}} \sim 10^{-5}$ s, the centre of mass of the crystal acquires opposite final velocities of magntitude $v \sim \frac{\mu_B  \frac{\partial B}{\partial x}}{m}\tau \sim  10^{-19}$ m s$^{-1}$ for the $|\uparrow\rangle$ and $|\downarrow\rangle$ components respectively. It is not worth exposing the crystal to the high magnetic field gradient much longer as this gradient, in addition to the Stern-Gerlach splitting, also creates a diamagnetic trap of a frequency $\omega \sim \sqrt{\frac{\chi_m}{\mu_0}}\frac{\partial B}{\partial x} \sim 10^5$ Hz in the $x$ direction, which reverses the directions of the opposite accelerations after a quarter period. After this, the $x$ gradient is switched off, and the mass is allowed to freely evolve for the exceptionally long time $\tau \sim 10^5$ s of our experiment in the diamagnetic trap in the vertical $z$ direction. Since there is no trapping/potential in the $x$ direction, the velocity difference is translated to a position difference $\Delta x \sim 2 v \tau \sim 10^{-14}$ m. Note that during the stages in which the Stern-Gerlach effect is not used to actively accelerate the crystal, the electronic spin states used in the Stern-Gerlach splitting can be mapped on to nuclear spin states which maintain their quantum coherence for exceptionally long times \cite{zhong2015optically}.

Another method to create a quantum superposition will be through using an optomechanical interaction with a quantized microwave field in a cavity, with the cavity field subsequently mapped to spin qubits. In this case, the optomechanical force is sufficient to create the required superposition of a single $m\sim 1$ g mass directly. An electromagnetic field in a number state $|n\rangle$ in a cavity interacts with a crystal passing through it with the coupling strength $g \sim \frac{3V}{4V_c}\frac{\epsilon-1}{\epsilon+1}\omega_L$, where $\omega_L$ is the frequency of the electromagnetic field, $V$ is the volume of the crystal, $V_c$ is the cavity waist volume and $\epsilon$ the dielectric constant \cite{chang2010cavity}. The optomechanical coupling Hamiltonian is $\hbar g k \hat{n} \hat{x}$, where $\hat{n}$ and $\hat{x}$ are the number and the position operators of the field and the centre of mass of the crystal respectively, and $k$ is the wavevector of the electromagnetic field. If interacting for a time $t_{\text{kick}}$, a crystal can receive a velocity kick $v_{\text{kick}} \sim \frac{\hbar g k n t_{\text{kick}}}{m}$. Assuming a $t_{\text{kick}}\sim 1 \mu$s during which the $1$ g mass traverses through the cavity waist, assuming both $V \sim V_c \sim 1$ cm$^3$ and $\omega_L \sim 10$ GHz, we get a velocity kick of $v_{\text{kick}} \sim 10^{-19}$ ms$^{-1}$. Thus, one can prepare the cavity in a quantum superposition $\frac{1}{\sqrt{2}}(|0\rangle_c+|1\rangle_c)$ and apply a kick to the crystal by letting it fall through the cavity for $t_{\text{kick}}$. After waiting for a time $\tau \sim 10^5$ s, one obtains a superposition $\frac{1}{\sqrt{2}}(|0\rangle_c|\bar{x}_0\rangle_{C}+|1\rangle_c|\bar{x}_1\rangle_{C})$ \cite{bose1999scheme}. The microwave cavity state can also be mapped to a long-lived nuclear spin states of trapped atoms after the state dependent velocity kicks are over.\\
Note that in all the above discussions, it is implicit that the interferometry has to be completed. So further spin or cavity field dependent impulses will be required at certain points to stop the growth of $\Delta x$ and reverse it, as it is accomplished in various interferometric schemes \cite{bose_spin_2017,marshman_mesoscopic_2020}.\\
Note that the phase growth between the components of the superposition when subject to a neutrino flux is linear with time at a constant rate of $\sim 10^{-5}$ Hz, which reflects in population differences as seen form Fig.\ref{fig:PhaseAccumulation}. At $\sim 10^5$ s, the phase difference of the order $\pi$ is obtained so that the final state of the superposition is orthogonal and therefore fully distinguishable from the initial state. If we can further ensure that, by resorting to means such as those described in the context of temperature and pressure, variations of the phase with the location and direction relative to the neutrino source as well as the use of different materials, {\em only} neutrinos have been scattered during this duration, then an orthogonal spin state detection at $\sim 10^5$ s corresponds to a single neutrino detection (a click), as we are ensuring that nothing else causes a change in the phase and the chance of more than one neutrino having interacted with the gram scale detector is vanishingly small. Using an array of $10^4$ such gram scale detectors, we should be able to ensure that one neutrino is detected every $10$ s. \\
Note also that one can shorten both the duration of the experiment to $\tau/n$ and the mass required to $m/n$ while keeping the detectability of the phase effect at the same level by using $n^4$ crystals. Each such crystal will get $n$ times the velocity kick of a single crystal of mass $m$ for the same $t_{\text{acc}}$ or $t_{\text{kick}}$ (we assume $\tau/n >> t_{\text{acc}}, t_{\text{kick}}$). Thus the time-scale of generation of a superposition of given size $\Delta x$ will become shortened to $\tau/n$. Note that the phase accrued by {\em each} crystal in this shorter time will decrease by a factor of $n^2$, i.e., $\phi$ becomes $\phi/n^2$ as both the time and the mass to which the phase is directly proportional, decrease by $n$. Because of shot noise scaling $n^4$ interferometers can measure a phase of $\phi/n^2$ with the same accuracy by measuring the spins of each interferometer.

\section{Challenges}
We have discussed the conditions needed to meet one of the principal challenges, namely environment induced decoherence, in the previous section while discussing the generation of the superposition. However, we discuss below how some of the other requirements may be met.
\subsection{Satisfying the Requirements of the Crystal Wavepacket}
Note that the initial spread $\sigma_c$ of each of the superposed Gaussian wavepackets of the crystal are required to satisfy a couple of conditions in order to meet some of the simplifying approximations of our calculations. Note that the position degree of freedom of all the nuclei (being part of a solid) are, to a good approximation (at least at the temperatures we consider) rigidly tied to the centre of mass of the crystal and thereby has the same position spread $\sigma_c$ as that of the whole crystal. As stated in section \ref{sec:two}, we require $\sigma_c \lesssim 1/q_x$, relating to the momentum transferred, which in this case boils down to ensuring $\sigma_c \lesssim \Delta x \sim 10^{-14}$ m. On the other hand, we also require the initial maximum momenta of the crystal $k_i \sim 1/\sigma_c \lesssim m_{\text{nucl}} \sim 10^{-17}$ m, which stems from the assumption of the heavy nucleus being effectively at rest. Thus there is a window. An initial diamagnetic trap of $10^5$ Hz in the $x$ direction will thus do the job, with the ground state spread of the centre of mass of the crystal in such a trap being $\sim 10^{-16}$ m.

\subsection{Coherence Length of the Neutrino}
The consideration of processes involving neutrinos brings about several unknowns, one of which is the particle's coherence length. In ref~\cite{chan_wave-packet_2016} the authors opted for a wave-packet treatment of the neutrino and estimated that an energy uncertainty of $\sigma_{wp}=\frac{\sigma_\nu}{E(p_\nu)}\sim0.01$ or larger would influence decoherence and dispersion effects and thereby reduce the detector efficiency of reactor anti-neutrino oscillation experiments. We take this value as a reference to estimate whether a scattering event could resolve the position of our nucleus and hence spoil the superposition. Considering $\sigma_x\sigma_\nu\sim\frac{\hbar}{2}$, we obtain an uncertainty $\sigma_x\sim 3\cdot10^{-12}$~\si{m} for a neutrino with energy $E_\nu\sim 10~\si{MeV}$. This means that a neutrino with $\sigma_{wp}\sim0.01$ would indeed be able to resolve the position of any quantum object in a superposition larger than $\sigma_x$. Seeing as the matter of the actual wave packet shape and coherence length of the neutrino is not solved, our proposed experiment may also be able to serve as a means of testing the validity of plane wave approximations of the neutrino wave packet. Depending on the true wave packet shape of the neutrino, we expect to observe either a coherent phase gain or a decoherence effect.

\subsection{Lattice Defects}

The authors of \cite{rajendran_method_2017} considered the structural damage effects of dark matter and neutrino scattering on dense materials with defect centers, such as nitrogen vacancy centers in diamond. In general, the deposited kinetic energies will exceed typical lattice binding energies of $\mathcal{O}(10)\si{eV}$. Hence, we anticipate the scattering of a neutrino to lead to the formation of such damage clusters, though they can be expected to be significantly smaller in size. Based on the analysis in~\cite{rajendran_method_2017}, we estimate that a nucleus recoiling with an average kinetic energy of 3$\si{keV}$ could generate $\mathcal{O}(10)$ lattice defects or interstitial sites. We require that these sites are created in such a manner that we cannot tell from which part of the superposition the neutrino has scattered. Here this requirement is naturally fulfilled as the size of the superposition is much smaller than the interatomic spacing in the lattice.

\subsection{Distribution of Momentum}
It is the centre of mass of the whole crystal which is coupled to the embedded spin used for the superposition creation/recombination and the phase read-out. 
However, the neutrino is initially going to impart its energy to one of the nuclei in the whole crystal. At this stage, the energy imparted is localized to this nucleus, but the centre of mass already has the imparted momentum; however, the crystal cannot be considered as all rigidly connected nuclei moving together, which is required for the embedded spin to sense the transferred momentum. A local phonon is excited in the crystal at the site of the scattering neutrino. However, phonon relaxation times in crystals are generally $\sim 1 - 100$ ns \cite{zhao2008full}, after which the energy would have been transferred to the centre of mass of the whole crystal.

\section{Discussion}
We have described the detection of neutrinos from the relative phase they impart between the components of a quantum superposition of two spatially localized states of the centre of mass of a crystal. As naturally there is a distribution of momentum after the scattering, this process also causes a decoherence in addition to imparting a change in the relative phase. We thus formulated a master equation technique to evolve the full density matrix of the COM of a crystal under the scenario of the scattering of a relativistic particle from it. Solving that, we found that the optimal detection requires a $\sim 1$ g mass placed in a quantum superposition of states separated by a distance $\Delta x \sim 10^{-14}$ m. For completeness, we have also suggested a schematic and parameter domain by adopting which such superpositions could be achieved and mainatined for the long duration of our experiment, although much more analysis will be required for realistic scenarios. 

It is worth clarifying the role of the various ingredients of our method proposed herewith. The superposition serves as a means to detect the momentum recoil $k$ of the crystal in terms of the relative phase $k \Delta x$. Of course, that will happen for a crystal of any mass, including individual atoms in a superposition of states separated by $\Delta x$. However, in that case the cross section is very small. For a crystal of $N_{\text{Atoms}}$, the cross section is amplified $N_{\text{Atoms}}$ times. While an uncorrelated collection of $N_{\text{Atoms}}$ in the same superposition state will have the same cross section, the neutrino will only scatter from one of those atoms, and one has to measure each atom after appropriate basis rotations in order to measure whether one of them obtained a phase. In the case of a crystal with a single embedded spin, the phase gained by the neutrino hitting any one of the nuclei is mapped on to the relative phase between the COM states of the whole crystal. This is because of the very strong correlations between the positions of the atoms 
the $N_{\text{Atoms}}$, since they are all either clustered around the state $|\bar{x}_0\rangle$ or the state $|\bar{x}_1\rangle$. The positions of all the atoms are entangled when the centre of mass is placed in the superposition $|\bar{x}_0\rangle + |\bar{x}_1\rangle$. Moreover, due to the very mechanism of our interferometry, the embedded spin or other ancillary system is also entangled with the centre of mass during the interferometry, so that at the end, the phase can be estimated exclusively by measuring this single spin. The reason that we have used a regime in which the scattering from each nulceus is coherent is because it enhances the cross section by $N^2$ times ($N$ being the number of neutrons in a nucleus), which makes our times-scales about $10^4$ times less than what it would have been otherwise. 

It is important to clarify that going to either higher or lower energy neutrinos is not a trivial problem. For higher energies, it is true that the cross section increases as $\propto E^2$, where $E$ is the energy of the incident neutrinos. However, the momentum transferred may be too high and knock a nucleus completely out of the crystal so that the momentum is not imparted to the rest of the crystal. Moreover, $k\Delta x  \sim 1$ implies that much smaller superpositions $\Delta x$ will have to be used, which implies a great difficulty in satisfying our simplification assumptions $1/m_{\text{nucl}} << \sigma_c << \Delta x \sim 1/k$ and the calculations will be needed to be performed in much more generality. For lower energies, cross section can both decrease due to the $\propto E^2$ effect or increase due to scope of scattering coherently from all atoms of the crystal. However, producing a larger $\Delta x \sim 1/k$ superposition becomes much more difficult, especially for the masses as large as the ones that are needed here.   

Our technique presented in this work should be adaptable to any relativistic particles scattering off a quantum superposition with appropriate modifications. Moreover, neutrinos will form a background to any other signals one may want to detect. The calculations here show that even a substantially large mass in a quantum superposition of distinct spatial states can remain coherent for a very long time close to a source of neutrinos so that there is an ample window for the detection of other signals.

\section{Acknowledgements}

EK acknowledges support from the Engineering and Physical Sciences Research Council [grant number EP/L015242/1]. MT and SB would like to acknowledge EPSRC grant No.EP/N031105/1, SB the EPSRC grant EP/S000267/1, and MT funding by the Leverhulme Trust (RPG-2020- 197). FFD acknowledges support from a UK STFC consolidated grant (Reference ST/P00072X/1). RS would like to acknowledge support from the Science and Technology Facilities Council [grant number ST/S000666/1].

\twocolumngrid

\bibliographystyle{apsrev4-2} 

\bibliography{references.bib}

\end{document}